%% file: main.tex
\documentclass[%
 reprint,
 amsmath,amssymb,
 aps,
]{revtex4-2}
\usepackage{graphicx} 

\usepackage{siunitx}
\usepackage{upgreek}

\input{my_math_definitions.tex}

\begin{document}

\preprint{APS/123-QED}

\author{J.N. K\"ammerer}
\affiliation{Physikalisches Institut, Karlsruher Institut für Technologie}
\author{S. Masis}
\affiliation{Physikalisches Institut, Karlsruher Institut für Technologie}
\author{K. Hambardzumyan}
\affiliation{Physikalisches Institut, Karlsruher Institut für Technologie}
\author{P. Lenhard}
\affiliation{Physikalisches Institut, Karlsruher Institut für Technologie}
\author{U. Strobel}
\affiliation{Physikalisches Institut, Karlsruher Institut für Technologie}
\author{J. Lisenfeld}
\affiliation{Physikalisches Institut, Karlsruher Institut für Technologie}
\author{H. Rotzinger}
\affiliation{Physikalisches Institut, Karlsruher Institut für Technologie}
\affiliation{Institut für Quantenmaterialien und Technologie, Karlsruher Institut für Technologie}
\author{A.V. Ustinov}
\affiliation{Physikalisches Institut, Karlsruher Institut für Technologie}
\affiliation{Institut für Quantenmaterialien und Technologie, Karlsruher Institut für Technologie}

\title{Resonant escape in Josephson tunnel junctions under millimeter-wave irradiation}

\begin{abstract}
The microwave-driven dynamics of the superconducting phase difference across a Josephson junction is now widely employed in superconducting qubits and quantum circuits. With the typical energy level separation frequency of several GHz, cooling these quantum devices to the ground state requires temperatures below 100 mK. Pushing the operation frequency of superconducting qubits up may allow for operation of superconducting qubits at \SI{1}{\kelvin} and even higher temperatures. Here we present measurements of the switching currents of niobium/aluminum-aluminum oxide/niobium Josephson junctions in the presence of millimeter-wave radiation at frequencies above \SI{100}{\giga\hertz}. The observed switching current distributions display clear double-peak structures, which result from the resonant escape of the Josephson phase from a stationary state. We show that the data can be well explained by the strong-driving model including the irradiation-induced suppression of the potential barrier. While still being measured in the quasi-classical regime, our results point towards a feasibility of operating phase qubits around \SI{100}{\giga\hertz}.
\end{abstract}

\keywords{Josephson junction, phase escape, mm-waves, quantum circuits}
\maketitle

\section{Introduction}

Quantum circuits employing superconducting qubits have seen a lot of progress over the past two decades~\cite{kjaergaard_superconducting_2020}. These circuits operate at microwave frequencies, typically around \SIrange[range-phrase = \,-\, ,range-units=single]{4}{8}{\giga\hertz}, and temperatures around \SI{20}{\milli\kelvin}. Cooling is required to reduce the population of the excited states of the qubits, which translates into fulfilling the temperature condition $T\ll T_0=hf/k_B$. Here $h$ is Planck's constant, $f$ is the frequency of the transition between the ground and first excited state of the qubit, and $k_B$ is Boltzmann's constant. For the transition frequency $f=\SI{4}{\giga\hertz}$, the temperature $T_0$ is about \SI{200}{\milli\kelvin}. However, it appears interesting and appealing trying to increase the qubit operation frequency to explore the possibility of operating them at temperatures much higher than currently required for superconducting quantum computers.

Increasing the qubit operation frequency beyond \SIrange[range-phrase = \,-\,,range-units=single]{20}{30}{\giga\hertz} is hardly possible when using aluminum, which is currently the standard material for most modern quantum circuits. Aluminum has the superconducting gap energy on the order of \SI{0.18}{\milli\electronvolt}~\cite{roberts_survey_1976}, which is equal to the energy of photons with a frequency of \SI{87}{\giga\hertz}. An appropriate superconducting material for making high-frequency qubits can be niobium or niobium nitride, both having superconducting gap energies larger by an order of magnitude than aluminum. Furthermore, reaching the above goal also requires junctions with significantly higher Josephson plasma frequency, which primarily depends on the critical current density of the junctions. 

In this work, we report millimeter(mm)-wave spectroscopy of the potential well profile  for a current-biased \nbaloxnb{} junction. Similarly to the previously studied in the microwave range superconducting phase qubit~\cite{martinis_energy-level_1985,martinis_experimental_1987,wallraff_multiphoton_2003}, as a function of the bias current we measure the statistics of the escape events from the superconducting state by applying electromagnetic radiation in the \SIrange[range-phrase = \,-\,,range-units=single]{100}{110}{\giga\hertz} frequency range. At the chosen mm-wave radiation power level, the switching current distributions display double-peak features, which result from the resonant escape from a stationary state~\cite{fistul_quantum_2003}. While the measurements here are performed at relatively high temperature of \SI{4.2}{\kelvin} in the regime of the thermally-activated escape, our results validate the successful fabrication of low-loss Josephson tunnel junctions with high plasma frequency and demonstrate the possibility of operating Josephson phase qubits at mm-wave frequencies.

\section{Theory}
A description of the Josephson phase dynamics including the behavior at finite voltage is provided by the resistively and capacitively shunted junction (RCSJ) model~\cite{tinkham_introduction_2004}. This model takes into account the dissipative quasiparticle current at finite voltage $V$ across the Josephson junction by means of the 
linear shunt resistance $R$.
The shunt capacitance $C$ models the geometric capacitance of the junction electrodes.
The effect of thermal fluctuations is taken into account by the addition of a Johnson noise current \fluccurrent{} which is determined by the shunt resistance $R$ at temperature $T$~\cite{ambegaokar_voltage_1969,fulton_lifetime_1974,martinis_experimental_1987}.
The dynamical behavior of the junction is described by~\cite{martinis_experimental_1987}
\begin{equation}\label{eq:Theory:RCSJ_phase}
    C\left(\frac{\fluxquant}{2\mypi} \right)^2\ddot{\phase} + \frac{1}{R}\left(\frac{\fluxquant}{2\mypi} \right)^2\dot{\phase} + \frac{\critcurrent\fluxquant}{2\mypi}\left(\sin\phase - I/\critcurrent \right)  + \fluccurrent (t) = 0\,,
\end{equation}
where $I$ is the total current flowing through the junction, \critcurrent{} denotes the critical current of the junction, \fluxquant{} represents the flux quantum and \phase{} is the Josephson phase, i.e., the phase difference across the junction.\par
Eq.~\eqref{eq:Theory:RCSJ_phase} has the mechanical analog of a particle of mass $m = (\fluxquant/ 2\mypi)^2 C$ moving along a tilted-washboard potential~\cite{tinkham_introduction_2004} 
\begin{equation}\label{eq:Theory:RCSJ_potential}
U(\phase) = -\josen\cos\phase - \left(\frac{\fluxquant I}{2\mypi}\right)\phase \,,
\end{equation}
where the phase \phase{} is interpreted as the position and $\josen = \critcurrent\fluxquant/(2\mypi)$ denotes the Josephson energy. 
The curvature at the potential minima $\phase_m$ of \eqref{eq:Theory:RCSJ_potential} defines the frequency of small oscillations~\cite{fulton_lifetime_1974}
\begin{equation}\label{eq:Theory:RCSJ_omso}
\omso (I) = \ompl\cdot\left[ 1 - \left(\frac{I}{\critcurrent} \right)^2 \right]^{1/4}\,,
\end{equation}
with $\ompl = (2\elemcharge\critcurrent / \hbar C)^{1/2}$ the plasma frequency of the junction.
The quality factor $Q = \omso R C\,$ describes the damping of the classical oscillatory behavior of the phase at the small oscillation frequency \omso{}~\cite{martinis_experimental_1987}.
When the bias current surpasses \critcurrent{}, the minima $\phase_m$ vanish and the particle is moving down the potential hill, which causes a finite voltage $V$ across the junction.\par
For bias currents close to the critical current $(1 - I/\critcurrent)\ll 1$, the wells of the tilted-washboard potential are well approximated by a cubic potential with barrier height $\Delta U=(4\sqrt{2}\josen/3)(1-I/\critcurrent)^{3/2}$~\cite{fulton_lifetime_1974,martinis_experimental_1987}.
\subsection{Escape in the thermal regime}
In underdamped junctions, defined in terms of the McCumber parameter $\mccumber = (\ompl RC )^2$ as junctions with $\mccumber \gg 1$, thermal noise enables switching of the junction from the superconducting to the voltage state for bias currents $\swcurrent < \critcurrent$~\cite{fulton_lifetime_1974}.  This process is described by a thermally activated escape of the phase from a metastable well. Following the calculations conducted by Kramers~\cite{kramers_brownian_1940}, the thermal escape rate is given by~\cite{weiss_quantum_1999,martinis_experimental_1987,grabert_quantum_1987}
\begin{equation}\label{eq:Theory:thermescrate}
    \thermescrate = a_t \frac{\omso}{2\mypi}\exp\left( -\frac{\Delta U}{\boltzm T} \right)\,,
\end{equation}
where $\Delta U$ denotes the barrier height and \omso{} plays the role of an attempt frequency. 
The damping-dependent transmission-factor $a_t$ in the regime of weak to moderate damping and in cubic approximation is given by $a_t = 4a/[(1+a Q\boltzm T/1.8\Delta U)^{1/2} + 1]^2$ with a numerical constant $a \approx 1$~\cite{buttiker_thermal_1983}.\par
The crossover temperature, given by $\tcross = \hbar\omso/(2\mypi\boltzm) $ in the weak-damping limit, separates the thermal regime $T\gg\tcross$ from the quantum regime $T\ll\tcross$, where the escape is dominated by macroscopic quantum tunneling~\cite{grabert_crossover_1984,martinis_experimental_1987,grabert_quantum_1987}.\par
The escape temperature \tesc{} is related to the escape rate $\Gamma$ via~\cite{martinis_experimental_1987} 
\begin{equation}\label{eq:Theory:escrate_esctemp}
\Gamma = \frac{\omso}{2\mypi}\exp\left( -\frac{\Delta U}{\boltzm \tesc }\right)\,.
\end{equation}
To a good approximation, \tesc{} is independent of the bias current and provides a measure for the escape process in addition to the bias-current dependent $\Gamma$. 
In the thermal regime, the escape temperature is derived from Eq.~\eqref{eq:Theory:thermescrate} as
\begin{equation}
\tesc = \frac{T}{1 - \ln a_t / \left[ \Delta U / \left(\boltzm T\right) \right] }\,,
\end{equation}
which is expected to be close to the actual temperature~\cite{martinis_experimental_1987}.\par
Making use of the cubic approximation for the potential barrier, Eq.~\eqref{eq:Theory:escrate_esctemp} yields~\cite{fulton_lifetime_1974,martinis_experimental_1987}
\begin{equation}\label{eq:Methods:sw_curr_thermregime}
\Gamma = \frac{\omso}{2\mypi}\cdot\exp\left\{ \left( \frac{\josen}{\boltzm \tesc}\frac{4\sqrt{2}}{3} \right) \cdot \left(1- \frac{I}{\critcurrent}\right)^{3/2}\right\}\,.
\end{equation}
Therefore, the escape rate calculated from the measured switching current statistics in the thermal regime depends on three parameters: \critcurrent{}, \tesc{} and \ompl{}.
\subsection{Escape under external driving}\label{sec:theory:resonant_escape}
The irradiation of microwaves onto a Josephson junction causes a resonant enhancement of the escape rate~\cite{martinis_energy-level_1985,martinis_experimental_1987,wallraff_multiphoton_2003,fistul_quantum_2003,gronbech-jensen_microwave-induced_2004,yu_resonant_2013}, which has been observed in measurements of the switching current distribution below~\cite{martinis_energy-level_1985,wallraff_multiphoton_2003} as well as above the crossover temperature $\tcross$~\cite{gronbech-jensen_microwave-induced_2004}. This microwave-induced rate enhancement results in a multi-peaked switching current distribution and the resonant peak position is dependent on the irradiation frequency, in good agreement to the frequency of small oscillations~$\omso{}(I)$~\cite{wallraff_multiphoton_2003,gronbech-jensen_microwave-induced_2004}.\par
A quantum model for the escape in the strong-driving limit~\cite{fistul_quantum_2003} describes both the quantum and the thermal regime. In the presence of microwave irradiation at the frequency $\omega/(2\mypi)$ and power $P$, the escape mechanism of the phase in the limit of strong driving, defined by $(\omega/\ompl)^5\gg\hbar\ompl/\josen$, is governed by a process of effective barrier suppression. By taking into account thermal fluctuations, the average shift in switching current $\langle \delta \swcurrent (P) \rangle = 1 -  \langle\swcurrent (P)\rangle/\critcurrent$ is described by a transcendent equation
\begin{eqnarray}\label{eq:Theory:swdependence_power_thermal}
\langle \delta \swcurrent (P) \rangle &=& \langle \delta \swcurrent (0) \rangle + k^{-1} P\nonumber\\ &\times& \sum_{nm} \frac{f_{nm}^4}{\left[ \hbar^{-1} E_{nm}(\langle \delta \swcurrent (P) \rangle ) - \omega \right]^2 + \alpha^2}\,.\ 
\end{eqnarray}
Here, $k$ is the microwave coupling coefficient, $\alpha = \omega/\sqrt{Q}$ is the damping parameter, $E_{nm}(I) = E_n (I) - E_m (I) $ is the separation of the resonantly interacting energy levels and $f_{nm} = \langle n|\phaseop | m\rangle$ denotes the matrix elements of the phase operator. The damping parameter $\alpha$ does not only depend on the quasiparticle conductance but is also influenced by the frequency-dependent impedance of the junction leads~\cite{martinis_experimental_1987}.
For temperatures above \tcross{}, the fluctuation induced shift in a current sweep experiment with constant sweep-rate $\timediff{I}$ is given by
\begin{equation}
\langle \delta \swcurrent (0) \rangle \approx \left[\frac{\boltzm T}{2\josen}\cdot \ln \left(\frac{\ompl\critcurrent}{2\mypi\timediff{I}}\right) \right]^{2/3} \,. 
\end{equation}
Depending on damping, temperature and microwave power, Eq.~\eqref{eq:Theory:swdependence_power_thermal} possesses multiple solutions, explaining the existence of double-peak structures in the switching current distribution~\cite{fistul_quantum_2003}.\par
In the harmonic approximation, the sum in Eq.~\eqref{eq:Theory:swdependence_power_thermal} is reduced to the term with $n = 0 ,\, m = 1$ and the transition frequency is approximated by the small oscillation frequency \omso{}~\cite{fistul_quantum_2003,yu_resonant_2013}. By utilizing the transition matrix element of the harmonic oscillator~\cite{martinis_decoherence_2003}, Eq.~\eqref{eq:Theory:swdependence_power_thermal} simplifies to~\cite{yu_resonant_2013} 
\begin{eqnarray}\label{eq:Theory:swdependence_power_harmonic}
\langle \delta \swcurrent (P) \rangle &=& \langle \delta \swcurrent (0) \rangle + k^{-1} P\frac{4\elemcharge^4}{\hbar^2 \omso^2C^2} \nonumber\\ &\times&\left( \left[ \omso (\langle \delta \swcurrent (P) \rangle ) - \omega \right]^2 + \frac{\omega^2}{Q}\right)^{-1}\,.
\end{eqnarray}
\section{Methods}
\begin{figure}
\includegraphics{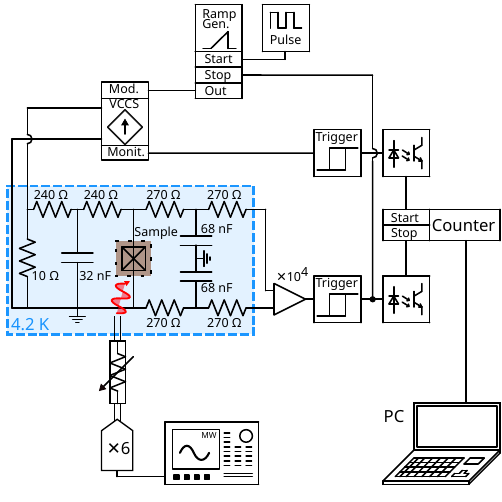}
\caption{\label{fig:Setup}Schematic of the switching current measurement setup. A time interval counter determines the duration between the zero-crossing of the bias current ramp and the moment a finite voltage develops across the junction, exceeding a threshold value. The knowledge of the current-ramp rate allows to convert the resulting time interval into a switching current. The setup allows for the irradiation of the sample by mm-waves in the frequency range from \SIrange{75}{110}{\giga\hertz}.}
\end{figure}
The investigated samples were fabricated with a  \nbaloxnb{} trilayer process, as described elsewhere~\cite{strobel_notitle_nodate}. The expected critical current density is in the range of  
\SI{2000} to \SI{2200} {\ampere\per\square\cm} 
at \SI{100}{\milli\kelvin}, while the expected junction capacitance is around \SI{80}{\femto\farad\per\square\micro\meter}. This set of fabrication parameters aims for a junction plasma frequency at zero bias around $\ompl/2\mypi = \SI{140}{\giga\hertz}$. The nominal junction area is $\SI{4}{\micro\meter}\times\SI{4}{\micro\meter}$.
\subsection{Measurement setup}
The sample is measured in liquid helium at \SI{4.2}{\kelvin} by means of a dipstick cryostat containing a dielectric waveguide.
The dielectric waveguide is connected to a horn to which the sample holder is attached, allowing for the irradiation of the sample by mm-waves.
Low pass and current divider filters submerged in liquid helium reduce the thermal noise from the room temperature setup.\par 
The measurement setup is schematically shown in Fig.~\ref{fig:Setup}. A current ramp with constant ramp rate $\dot{I}$ is generated by a sawtooth generator attached to a voltage-controlled current source at room temperature.
A pulse generator connected to the sawtooth generator restarts the ramp in an adjustable time interval $\Delta t_\mathrm{puls}$ and the zero-crossing of the current ramp triggers a start signal.
The voltage across the junction is amplified and a stop signal is triggered as soon as a threshold is exceeded. This stop signal terminates the current ramp. The interval $\Delta t_\mathrm{puls}$ is sufficiently large to ensure that the junction returns to the zero-voltage state.\par
The start and the stop signal are detected by a time interval counter \textit{SR620} from \textit{Stanford Research Systems} to measure the duration of the current ramp. The process is repeated continuously and the measured time intervals of the counter are read out by a computer. The time intervals are converted into switching currents through a calibration of the ramp rate.\par
The biasing electronic is electrically isolated from the digital data acquisition devices by optocouplers and optical fibers. Additionally, the part of the measurement setup located in liquid helium is protected by a mumetal shield.\par
Electromagnetic radiation in the W-band are generated by a low-noise frequency synthesizer \textit{APSYN140} from \textit{AnaPico} connected to an active multiplier \textit{QMC-MX6-10F10} from \textit{Quantum Microwave}, enabling the irradiation by mm-waves in the range from \SIrange{75}{110}{\giga\hertz} onto the sample. The power of the mm-waves is regulated with a voltage-variable attenuator \textit{VA100} from \textit{MicroHarmonics}. The attenuator is calibrated with the help of a vector network analyzer.
\subsection{Statistical analysis}
In the conducted current-ramp experiment, the junction is initially in the zero-voltage state and the bias-current is increased with a constant ramp rate~\cite{fulton_lifetime_1974}. The probability distribution $P_\mathrm{sw}(I)$ of the switching currents is defined such that the probability to measure a switching event in the small interval from $I$ to $I + \diff I$ is given by $P_\mathrm{sw}(I)\diff I$. Thus, repeatedly measuring \swcurrent{} allows to reconstruct $P_\mathrm{sw}(I)$ from which the escape rate is calculated.
The relation between $P_\mathrm{sw}(I)$ and the escape rate $\Gamma (I)$ is given by~\cite{fulton_lifetime_1974} 
\begin{equation}
P_\mathrm{sw}(I) = \Gamma (I) \left(\frac{\diff I}{\diff t}\right)^{-1} \left( 1 - \int_0^I P(\tilde{I})\diff \tilde{I} \right)\,.
\end{equation}
For the analysis of the experimental data, the measured switching current distribution is collected in a histogram with a bin width $\Delta I$ resulting in a discrete escape probability density~\cite{wallraff_fluxon_2001}
\begin{equation}
P_j = \frac{n_j}{N \cdot \Delta I}\,,
\end{equation} 
where $n_j$ is the number of counts in the $j$-th bin and N is the total number of counts. The according escape rate is~\cite{martinis_experimental_1987,wallraff_fluxon_2001}
\begin{equation}
\Gamma(I_k) = \frac{|\timediff{I}|}{\Delta I}\ln\left( \frac{\sum_{j \geq k} P_j}{\sum_{j\geq k+1}P_j } \right)\,.
\end{equation}
\section{Results}
\begin{figure}
\includegraphics{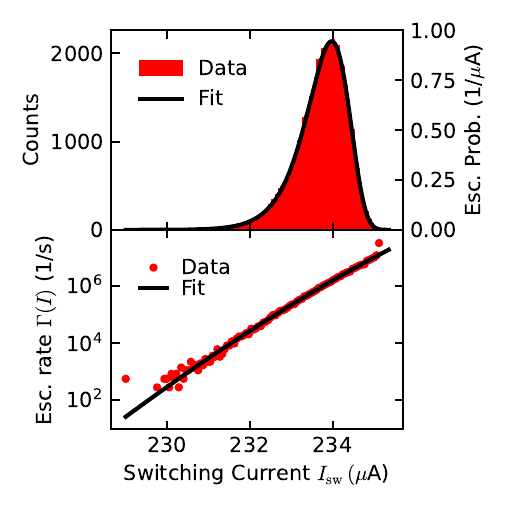}
\caption{\label{fig:swcurrent_thermal}\textbf{Top:} Switching current histogram in the thermal regime measured at \SI{4.2}{\kelvin}. 
\textbf{Bottom:} The escape rate calculated from the switching current distribution.
A fit of the calculated escape rate according to Eq.~\eqref{eq:Methods:sw_curr_thermregime} yields $\critcurrent = \SI{242\pm1}{\micro\ampere}$, $\tesc = \SI{6.3\pm0.4}{\kelvin}$ and $\ompl{}/(2\mypi) = \SI{0.1\pm0.2}{\tera\hertz}$. The result of the fit is represented by the black curve for both the escape rate and the escape probability. 
}
\end{figure}
\begin{figure}
\includegraphics{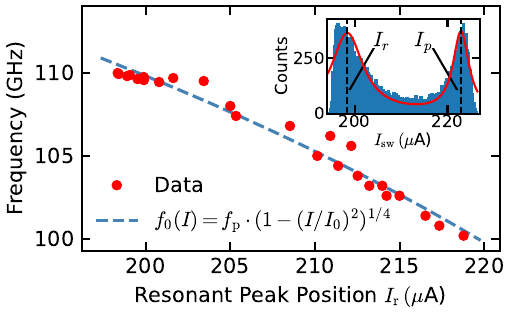}
\caption{\label{fig:swcurrent_spectroscopy}Results of the spectroscopic determination of the plasma frequency. The position of the resonant peak in the switching current distribution is measured as a function of the irradiation frequency.
The inset shows a switching current histogram measured under irradiation of mm-waves at \SI{109.92}{\giga\hertz} and a double-Lorentzian fit, performed in order to determine the peak positions. The dashed blue line shows a fit of the frequency of small oscillations $\freqso = \omso/(2\mypi)$ according to Eq.~\eqref{eq:Theory:RCSJ_omso} to the observed data with the parameters $\critcurrent = \SI{258\pm3}{\micro\ampere}$ and $\freqpl = \ompl/(2\uppi) = \SI{138.3\pm0.7}{\giga\hertz}$.}
\end{figure}
\begin{figure*}
\includegraphics{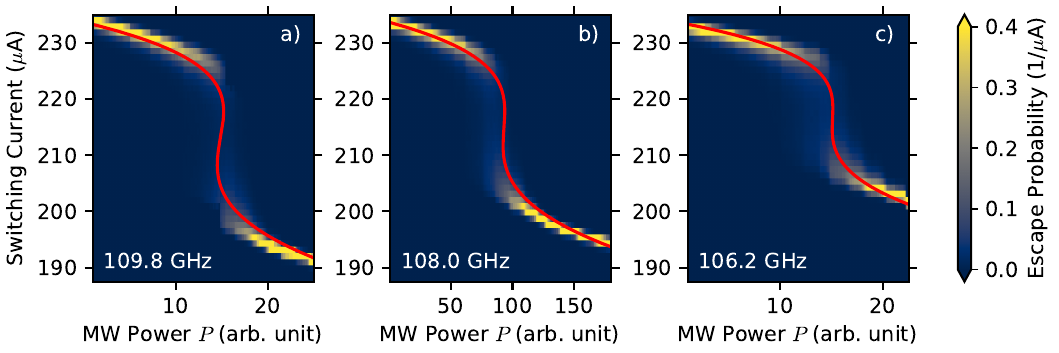}
\caption{\label{fig:powersweep_fits}Measurement of the switching current distribution in dependence of the applied mm-wave power for three different frequencies \textbf{a)} \SI{109.8}{\giga\hertz}, \textbf{b)} \SI{108.0}{\giga\hertz} and \textbf{c)} \SI{106.2}{\giga\hertz}. 
The red lines represent fits obtained from the solutions of Eq.~\eqref{eq:Theory:swdependence_power_thermal}. Fixed parameters are $\critcurrent = \SI{242}{\micro\ampere}$ and $T = \SI{4.2}{\kelvin}$. The plasma frequency $\ompl = 2\uppi\cdot \SI{145.5}{\giga\hertz}$ is a common fit parameter. The individual fit parameters are a) $Q \approx 97\,$,  $k^{-1}C^{-2} = \num{0.82e50}\,$(arb. unit), b) $Q \approx 91\,$, $k^{-1}C^{-2} = \num{0.115e50}\,$(arb. unit) and c) $Q \approx 90\,$, $k^{-1}C^{-2} = \num{0.6e50}\,$(arb. unit). }
\end{figure*}
First, the switching current distribution is measured without external mm-wave irradiation. The measurement is performed with a current-ramp rate of \SI{0.8}{\ampere\per\second} and a measurement frequency of \SI{200}{\hertz} for a total of \num{5e4} counts.
The result is presented in Fig.~\ref{fig:swcurrent_thermal}. The obtained asymmetric histogram with a bin width of $\Delta I \approx \SI{58.3}{\nano\ampere}$ possesses a mean switching current of $\langle \swcurrent \rangle \approx \SI{233.72}{\micro\ampere}$ with a standard deviation of $\sigma \approx \SI{0.62}{\micro\ampere}$. This yields a relative width of $\sigma / \langle \swcurrent \rangle \approx \num{2.7e-3}$.\par
A fit according to Eq.~\eqref{eq:Methods:sw_curr_thermregime} with \critcurrent{}, \tesc{} and \ompl{} as fit parameters is performed. In this least-squares fit, the weight of the data points corresponds to the square root of the number of escape events collected in each bin~\cite{martinis_experimental_1987}. This procedure does not allow for a precise determination of \ompl{}, but provides a value for the critical current $\critcurrent = \SI{242\pm1}{\micro\ampere}$.
The escape rates and escape probabilities calculated from the fit parameters are in agreement with the experimental data as shown in Fig.~\ref{fig:swcurrent_thermal}.\par
Under mm-wave irradiation, the switching current distribution, with the primary peak located at $I_p$, shifts to lower values with increasing mm-wave power, until a second resonant peak at position $I_r < I_p$ develops. This double-peak structure is shown in the inset of Fig.~\ref{fig:swcurrent_spectroscopy} for a radiation frequency of \SI{109.92}{\giga\hertz}. By increasing the mm-wave power further, the primary peak vanishes and the resonant peak shifts to lower values.\par
The frequency of the irradiated mm-waves is varied to measure the dependence of the resonant peak position $I_r$ on the radiation frequency. Here, the power of the irradiated mm-waves is adjusted, so that the primary and the resonant peak are of equal height. This ensures a statistically significant measurement of both peaks and provides a measure that allows to compare and reproduce the switching current measurements at varying frequencies. Each histogram is taken with at least \num{2e4} counts. The peak positions are determined by the fit of a curve with a double-Lorentzian shape for each histogram.\par
The results are depicted in Fig.~\ref{fig:swcurrent_spectroscopy}. A fit of $\freqso (I)= \omso/(2\mypi)$ according to Eq.~\eqref{eq:Theory:RCSJ_omso} to the observed data yields an estimate for \ompl{}. 
The accuracy of this heuristic approach is limited, as the resonant peak position $I_r$ is dependent on the mm-wave power. 
The plasma frequency $\ompl/(2\mypi) = \SI{138.3\pm0.7}{\giga\hertz}$, extracted as a fit parameter from this spectroscopic measurements, is close to the design expectations.
The corresponding maximum crossover temperature, evaluated at zero bias-current, is $\tcross = \SI{1.056\pm 0.005}{\kelvin}$, which is lower than the bath temperature of liquid helium \SI{4.2}{\kelvin}. This confirms that the measurements presented in this work are performed in the thermal regime.\par
Measurements of the switching current distribution in dependence of the irradiated mm-wave power are shown in Fig.~\ref{fig:powersweep_fits} for three different frequencies. For each mm-wave power, a measurement with a total number of \num{2e4} counts is taken.
It is visible how the branch of the primary peak is moving to lower switching currents for increasing mm-wave power. The development of a second resonant peak in coexistence with the primary peak is observed. By further increasing the mm-wave power the primary branch vanishes while the resonant branch becomes the main branch. In comparison, the jump between the primary and the resonant branch is less pronounced for lower frequencies. The relative width of the resonant peak, determined at a level of mm-wave power for which only the resonant peak is observed, is of the order of $\sigma / \langle \swcurrent \rangle \approx \num{1.8e-3}$.
This value is smaller than the relative width of the primary peak observed in the absence of mm-wave irradiation.\par 
Using $\critcurrent = \SI{242}{\micro\ampere}$, $\omega= 2\mypi\cdot\SI{100}{\giga\hertz}$ and $\ompl = 2\mypi\cdot\SI{138.3}{\giga\hertz}$ yields $\josen\cdot(\omega/\ompl)^5/\hbar\ompl \approx 170\gg 1\,$, i.e., the condition for the strong-driving limit is fulfilled. Therefore, the data for the power dependence is compared to the model of effective barrier suppression.
Differing widths for the primary and the resonant peak are explained by the theory of escape in the strong-driving limit, as the width of the primary peak is determined by thermal fluctuations while the width of the resonant peak depends on the damping of the junction~\cite{fistul_quantum_2003}.\par 
Figure~\ref{fig:powersweep_fits} also displays fits of the solutions of Eq.~\eqref{eq:Theory:swdependence_power_thermal} in the thermal regime to the measurement data. As fixed parameters, the critical current $\critcurrent{} = \SI{242}{\micro\ampere}$ obtained from the switching current distribution in the absence of mm-wave irradiation is used, as well as the bath temperature of liquid helium $T = \SI{4.2}{\kelvin}$. The fits for the three distinct frequencies of mm-wave radiation share the plasma frequency $\ompl = 2\mypi\cdot\SI{145.5}{\giga\hertz}$ as a common fit parameter. The obtained frequency-dependent effective quality factors are of the order of $Q \approx 90$, which we expect to be limited by quasiparticles at the temperature of \SI{4.2}{\kelvin}.\par
In summary, we studied the interaction of \nbaloxnb{} Josephson tunnel junction with mm-waves by investigating the switching current distribution. The analysis of the observed double-peak structures allows for the direct determination of the plasma frequency. 
Spectroscopic measurements of the dependence of the resonant peak bias current position on the radiation frequency are conducted.
Furthermore, the dependence of the switching current distribution on the mm-wave power at fixed frequency was analyzed by means of a quantum model in the strong-driving limit.
These measurements yield similar results and show that junctions with plasma frequencies of the order of $\ompl = 2\mypi\cdot\SI{140}{\giga\hertz}$ have successfully been fabricated.
The application of these techniques provides a quick turnaround methodology for the characterization of mm-wave Josephson junctions, allowing a rapid process development for qubit devices.

\section{Acknowledgments}

We thank T. Zwick for fruitful discussions. This work was supported by funding from the European Research Council (ERC) under the European Union's Horizon 2020 research and innovation programme (project {\em Milli-Q}, grant agreement number 101054327).

\bibliography{bibfile}
\end{document}

%% file: my_math_definitions.tex
\newcommand{\mypi}{\ensuremath{\uppi}}

\newcommand{\elemcharge}{\ensuremath{e}}
\newcommand{\fluxquant}{\ensuremath{\Phi_0}}

\newcommand{\boltzm}{\ensuremath{ k_{\mathrm{B}} } }

\newcommand{\nbaloxnb}{Nb/Al-AlO\textsubscript{x}/Nb}

\newcommand*\diff{\mathop{}\!\mathrm{d}}

\newcommand*\timediff[1]{\ensuremath{\dot{#1}}}

\newcommand{\fluccurrent}{\ensuremath{ I_\mathrm{F}}}

\newcommand{\phaseop}{\ensuremath{\hat{\delta}}}

\newcommand{\phase}{\ensuremath{{\delta}}}

\newcommand{\josen}{\ensuremath{ E_J }}

\newcommand{\mccumber}{\ensuremath{\beta_\mathrm{C}}}

\newcommand{\tcross}{\ensuremath{ T_{\mathrm{cr}} }}
\newcommand{\tesc}{\ensuremath{ T_{\mathrm{esc}} }}

\newcommand{\critcurrent}{\ensuremath{I_{\mathrm{c}} }}
\newcommand{\ompl}{\ensuremath{ \omega_{\mathrm{p}} }} 
\newcommand{\omso}{\ensuremath{ \omega_{0} }}
\newcommand{\freqso}{\ensuremath{ f_{0} }}
\newcommand{\freqpl}{\ensuremath{ f_{\mathrm{p}} }}

\newcommand{\swcurrent}{\ensuremath{I_\mathrm{sw}}}
\newcommand{\thermescrate}{\ensuremath{\Gamma_\mathrm{th}}}